\def\selectedoptions{final}

\documentclass[
   \selectedoptions
  ]
  {aipproc}

\layoutstyle{8x11single}

\SetInternalRegister\hbadness{8000}

\newcommand\doingARLO[2][]{%
  \ifx\mmref\undefined #1\else #2\fi
}

\begin{document}

\title [The Very Small Array] {The Very Small Array}


\author{Angela C. Taylor}{ address={Astrophysics Group, Cavendish
    Laboratory, Madingley Road, Cambridge, CB3 0HE, UK},
  email={act21@cam.ac.uk},
}

\copyrightyear {2001}

\begin{abstract}
  The Very Small Array (VSA) is a fourteen-element interferometer
  designed to study the cosmic microwave background on angular scales
  of 2.4 to 0.2 degrees (angular multipoles $l$ = 150 to 1800). It
  operates at frequencies between 26 and 36~GHz, with a bandwidth of
  1.5~GHz, and is situated at the Teide Observatory, Tenerife. The
  instrument also incorporates a single-baseline interferometer, with
  larger collecting area, operating simultaneously with and at the
  same frequency as the VSA main array. This provides accurate flux
  measurements of contaminating radio sources in the VSA observations.
  Since September 2000, the VSA has been making observations of
  primordial CMB fluctuations.  We describe the
  instrument, observing strategy and current status of the first year
  of observations.

\end{abstract}

\date{\today}

\maketitle

\section{Introduction}

\begin{table}[b]
\begin{tabular}{lrr}
\hline
&\tablehead{1}{l}{b}{Compact Array}
&\tablehead{1}{l}{b}{Extended Array}\\
\hline
Mirror size /mm & 143 & 322\\
Primary Beam (34~GHz) & 4.6$^\circ$  & 2.0$^\circ$\\
Synthesised Beam (34~GHz) & $\sim$30$^\prime$  & $\sim$11$^\prime$\\
$l$-range & 150-700 & 300-1800\\
$\Delta$S (28$\times$7~hr)~/~mJy beam$^{-1}$ & 30 & 6\\
$\Delta$T (28$\times$7~hr)~/~$\mu$K beam$^{-1}$ & 33 & 33\\
\hline
\end{tabular}
\caption{Basic VSA parameters}
\end{table}

The Very Small Array (VSA) is a 14-element interferometer designed to
make images of the cosmic microwave background (CMB) and to measure
its power spectrum over angular scales of 2.4 to 0.2 degrees ($l$ =
150 to 1800). The telescope is located at the Teide Observatory,
Tenerife at an altitude of 2400~m and, for observations in the region
26--36~GHz, the transparency at the site is approximately 98
percent. There is also negligible correlated emission from the atmosphere.

Interferometers are well-suited to measuring the power
spectrum of the CMB since they directly sample the Fourier modes on
the sky which can then be converted to a power spectrum. In addition
they provide excellent rejection of systematics since only correlated
signals are detected, reducing signals such as ground radiation and
atmospheric emission. Interferometric systems also offer the
opportunity to target a specific range of angular scales on the sky, determined by the spacing of the elements of the
array.

In order to achieve constant temperature sensitivity over the full
range of angular scales that the VSA is designed to measure, two
separate, but scaled, array configurations are used.  For measurement
of the CMB power spectrum over $l$-values 150--700, a `compact array'
configuration is used. Each receiver is fitted with a 143~mm diameter
antenna and typical baselines in this configuration range from
approximately 30~cm to 120~cm.  For mapping finer angular scales, the
same receivers are re-fitted with larger antennas, 322~mm in
diameter. The baselines of this `extended array' are scaled by the
same factor, thus allowing measurements to be made across the whole
range of $l$ with constant temperature sensitivity.
The specifications of these two arrays are given in Table 1.

The VSA project is a collaboration between the Cavendish
Astrophysics Group, Cambridge, Jodrell Bank Obsevatory, Manchester,
and the Instituto de Astrofisica de Canarias (IAC), Tenerife.
Although the telescope can be operated remotely, on-site support is
provided by the IAC. All three institutions are actively involved in
the data analysis.

\section{Design of the VSA}

\begin{figure}[!t]
\includegraphics[angle=-90]{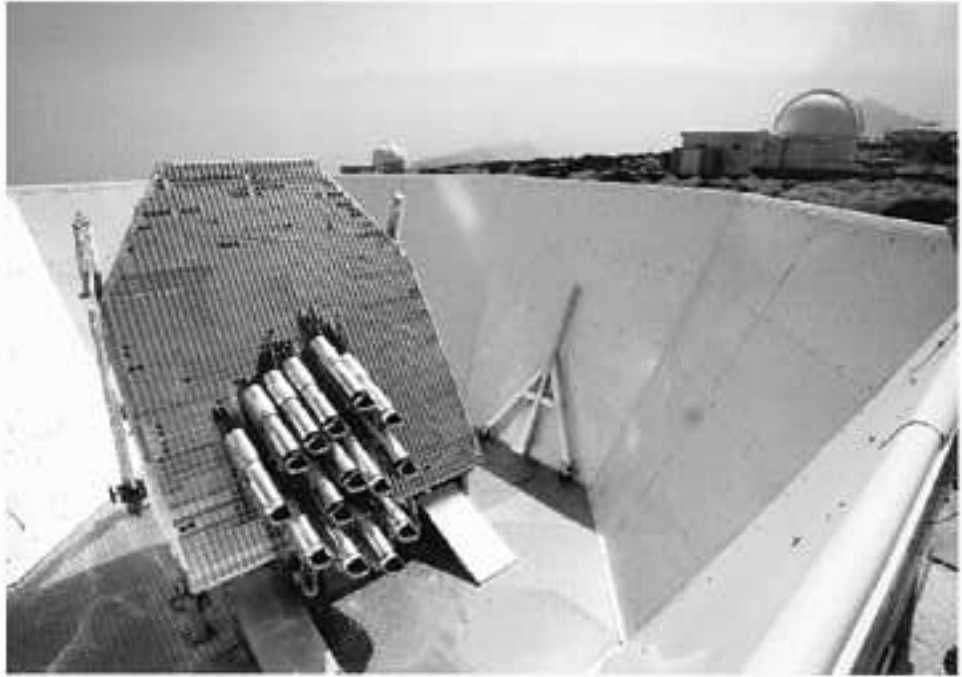}
\caption{The VSA main array in its enclosure.}
\end{figure}

The basic design of the VSA is a development of the Cosmic Anisotropy
Telescope (CAT) \cite{cat_inst}, a three-element interferometer which
operated at 15~GHz from a sea-level site in Cambridge. The VSA operates over the frequency range
26--36~GHz and has an observing bandwidth of 1.5~GHz.  The front-end
receivers are cryogenically cooled HEMT amplifiers and provide an
overall system temperature of 25--35~K.  Each receiver is
fitted with a corrugated horn-reflector antenna in which the mirror
can be rotated, providing tracking in one dimension. The fourteen
VSA receivers are all mounted on a tip-table, providing pointing
in a further dimension.  Close packing of the array is achieved by
mounting each element at 35$^\circ$ to the table.  The tracking range
of the table in elevation is 0$^\circ$--70$^\circ$ resulting in a range
of accessible declinations of $-7^\circ$ to $+63^\circ$. To
eliminate ground radiation, the complete array is surrounded by a a 3.5~m
high octagonal enclosure. Figure 1 shows the VSA main array in its
enclosure.

Quasi-independent tracking of the VSA antennas is a key feature
that distinguishes the VSA from other interferometric CMB experiments.
Since each antenna tracks individually, the astronomical signal path
to each antenna varies continuously during an observation.  The rate
of change of phase (or fringe rate) resulting from this continuous
change in path can be calculated for each baseline configuration. The
observed complex visibilities are then multiplied by the inverse of
this expected fringe rate to give a quasi-constant signal.  Since this
is equivalent to applying a matched filter, all signals not varying at
the expected fringe rate are removed.  This rejection of systematics
enables us not only to distinguish common systematics such as ground
radiation or residual cross-talk between antenna elements but also to filter out the effects of bright
sources such as the Sun and the Moon.  We have been successful in
removing the effects of both the Sun and Moon, even when they are as
close as 30$^\circ$ away from the VSA primary beam.  The
implementation and effectiveness of this filtering technique is
discussed further by M. Jones in these proceedings.

A building alongside the main enclosure houses the VSA
correlator and control room. The signal from each antenna is
down-converted to 8.25--9.75~GHz on the table, with further
down-conversion to baseband (0.25--1.75~GHz) before entering the
control building. Phase-switching is provided at the first stage of
down-conversion.  Once inside a screened room, path compensation is
provided, in increments of $\sim$~7~mm, by a sequence of strip-line
elements.  Appropriate pairs of signals are then fed by a series of
splitters to 182 correlators, providing the real and imaginary
components of the correlated signal from each of the 91 baselines.   
The outputs of the correlator are sampled every second.

In addition to the VSA main array we operate a separate
single-baseline interferometer, with much large collecting area,
adjacent to the main enclosure.  This 9-metre, north-south baseline
operates simultaneously with and at the same frequency as the VSA, and
forms part of our source-subtraction strategy described below.

\section{Source Subtraction}

\begin{figure}[!t]
\resizebox{.45\columnwidth}{!} {\includegraphics{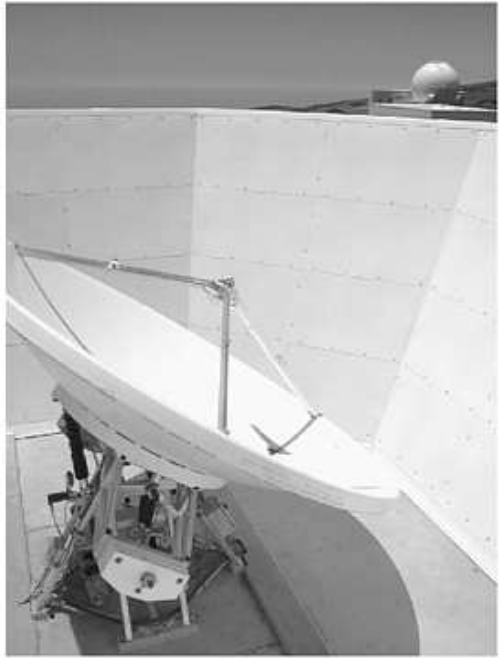}}
\caption{One of the 3.7m source-subtraction dishes.}
\end{figure}

Extragalactic radio sources are a major contaminant of CMB
observations at centimetre wavelengths.  Their population is not well
known at frequencies higher than 10~GHz, and many are expected to be
variable and/or have rising spectral indices, $\alpha$ (where S
$\propto \nu^{-\alpha}$).  On the angular scales of interest to
primordial CMB work, such sources are also generally unresolved. We
deal with this problem by observing all the contaminating radio
sources in each VSA field at higher resolution than can be reached
using the VSA. To achieve this, we implement a two-stage process.

First, prior to observation with the VSA, we survey all the VSA fields
at 15~GHz using the Ryle Telescope (RT) in Cambridge. The RT, which
uses five 13~m diameter antennas and gives a resolution of $\sim$30
arcsecs, is used in a raster scanning-mode and reaches an rms noise
level of $\sigma$ = 4mJy \cite{waldram_iau}. This allows us to
identify all sources above 20~mJy at 15~GHz, and ensures that we find all
sources above our source-confusion limit of 80~mJy at 34~GHz, even
allowing for a spectral index as steep as $-$2 between 15 and 34~GHz.

Having identified the contaminating sources in each field, we monitor
each source at 34~GHz using a separate single-baseline interferometer
working simultaneously with the VSA.  This single-baseline
interferometer consists of two 3.7m dishes separated by 9 metres on a
north-south baseline. Each dish is situated in an enclosure similar to
that of the VSA (Fig. 2.) and is fitted with identical horn-reflector
feeds and receivers as used on the main array. Every source identified
in the 15~GHz survey is monitored daily and its contribution is
subsequently subtracted in the $uv$ plane from VSA observations. The
monitoring is done simultaneously with and at the same frequency as
the VSA observation.  This ensures that sources which are
variable on time-scales as short as a few days can be subtracted
accurately.

\section{Observations}

\subsection{Field Selection}
\begin{figure}[!tl]
  \resizebox{.6\columnwidth}{!} {\includegraphics[angle=-90]{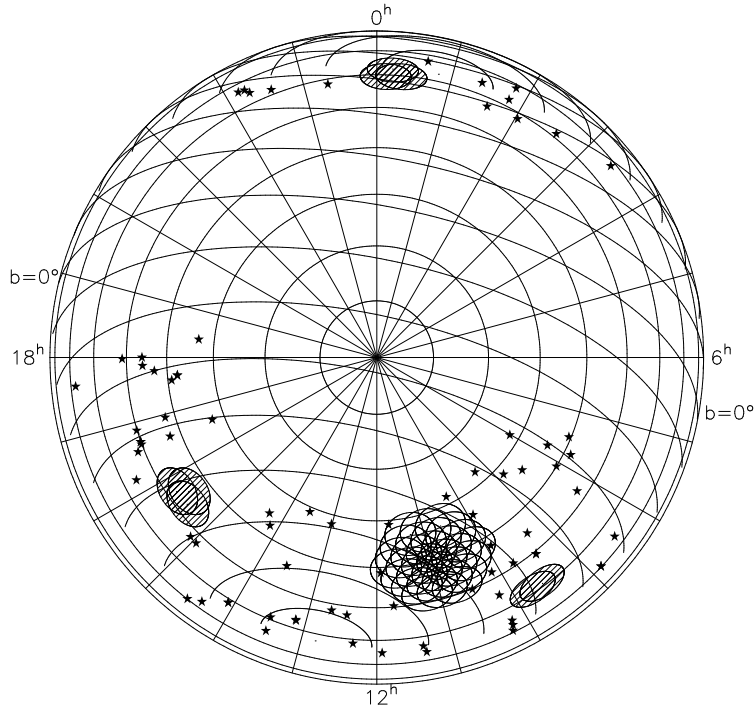}}
\caption{Plot of the VSA fields observed during our first year of
  observations.  The hatched regions are fields which were observed
  using deep mosaicing.  The remaining fields were observed as part of
  a shallow survey. Radio sources predicted to be brighter than
  500~mJy are displayed as star symbols.  A cut-off at galactic
  latitude b > 20$^{\circ}$ has been applied.}
\end{figure}
 
For practical CMB observation, it is important to choose fields which
are relatively free from Galactic and extragalactic foregrounds.  All
the VSA fields are situated at Galactic latitudes greater than
20$^\circ$ and have low Galactic synchrotron and free-free emission,
as predicted by the 408-MHz all-sky radio survey of Haslam et. al.
(1992){\citep{408_mhz}.  The dust maps of Finkbeiner et. al. (1994)
  {\citep{dust}} were used to select fields with relatively low dust
  contamination. To avoid large-scale structure and clusters we
  consulted the ROSAT catalogues {\citep{rosat},\citep{xbacs}}.  More
  importantly, all fields were chosen to be as free as possible of
  bright radio sources, since these are the major contaminant of CMB
  observations at 34~GHz.  We used two low-frequency surveys, NVSS
  {\citep{nvss}} at 1.4~GHz and Green Bank {\citep{gb6}} at 4.85~GHz
  to select CMB fields in which there are predicted to be no sources
  brighter than 500~mJy at 34~GHz.  Predictions were made by
  extrapolating the flux density of every source in the 4.8~GHz
  catalogue to 34~GHz on the basis of its spectral index between 1.4
  and 15~GHz. A further practical consideration which affected the
  choice of CMB fields was the need to observe all fields for a
  reasonable length of time, from both Tenerife and Cambridge.  This
  limited the declination range of our fields to +26$^\circ$~--~+54$^\circ$.  We
  also selected fields that are evenly spaced around the sky to
  enable 24-hour observing. In order to increase the $l$-resolution of
  our measurement of the CMB power spectrum, we selected regions of
  sky where we could mosaic several CMB fields. In each region of sky,
  mosaiced fields are separated by 2.75$^\circ$. Our final choice of
  fields used during the first year of observation is shown in Fig. 3.

\subsection{Observing strategy}

During the first year of VSA observations, we have undertaken two
distinct observation programs, each using a compact array.  First, we
have made deep mosaiced observations of eight fields in three evenly
spaced regions of sky (hatched regions in Fig. 3.).  Each mosaiced field
was observed for $\sim$~400 hours, reaching a thermal noise of
approximately 30~mJy.  Mosaicing in this way enables us to increase
the $l$-resolution of our measurements whilst also reducing sample
variance.  However, the time taken for us to survey all fields with
the Ryle Telescope prior to any observation with the VSA, has limited
the area of sky that we can cover with deep mosaicing in this first
year.  Consequently, for $l$$\le$~300, our measurements of the CMB
power spectrum are limited by sample variance.  In order to achieve a
good estimate of the CMB power spectrum in the region $l$$\le$~300, we
have now completed the second stage of our observation program; a
shallow survey in one area of sky.  The shallow survey consists of 2
days observation on each of 30 mosaiced fields (Fig. 3 hexagonal region), and covers an area of approximately 180 sq. degrees. For
this shallow survey, our source subtraction strategy no longer limits
the area of sky we can observe with the VSA in a given time.  Since we
are only concerned with low-$l$ observations, where the contribution
of point sources to the CMB power spectrum is known to be negligible,
the need for prior surveying with the Ryle Telescope is not necessary.
Instead, we choose only to monitor sources predicted to be greater
than 100~mJy at 34~GHz on the basis of low-frequency survey
information.
 
\begin{figure}
  \centering
  (a)\includegraphics[scale=0.4]{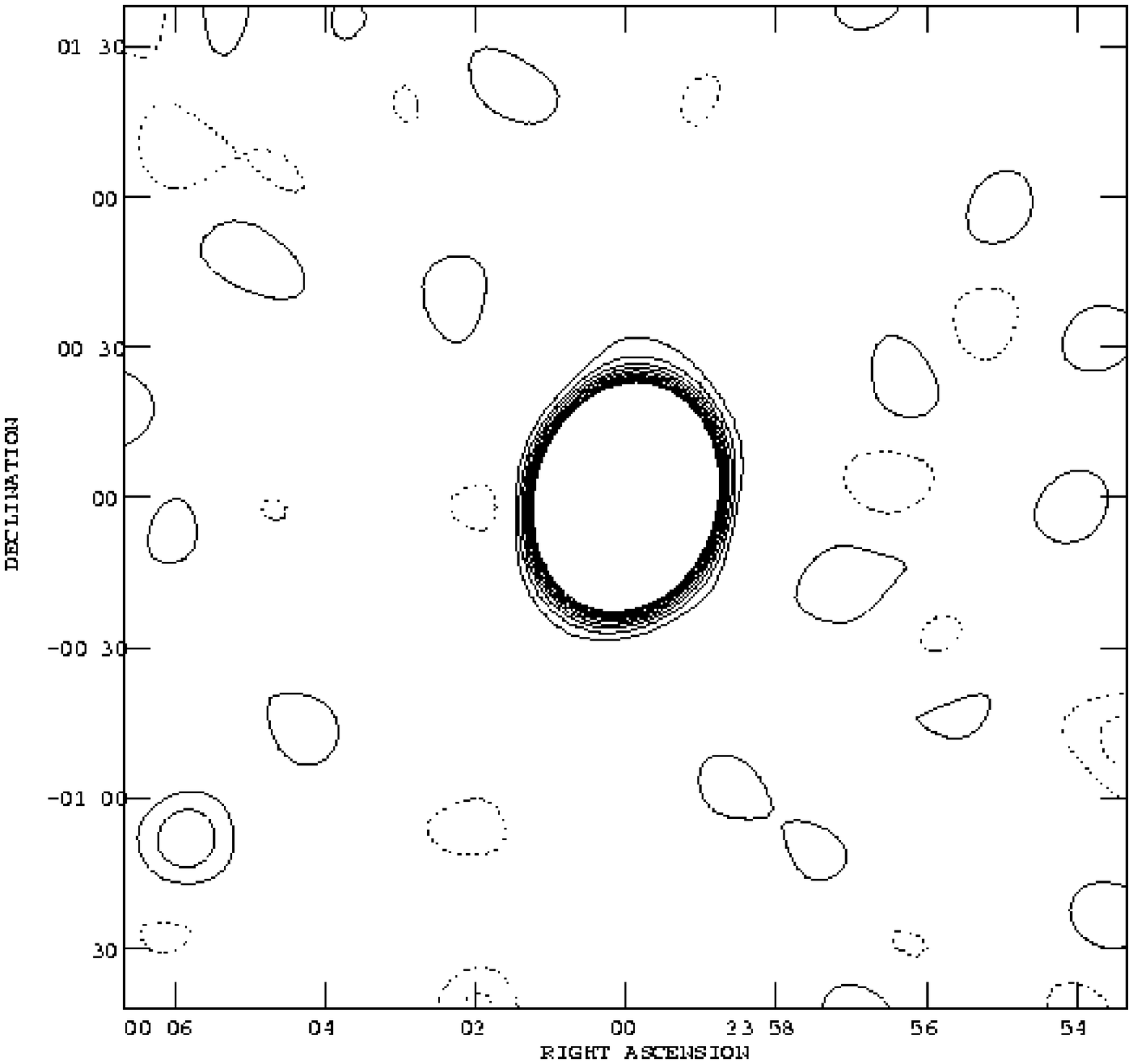}
  \hspace{0.7cm}
  (b)\includegraphics[scale=0.38]{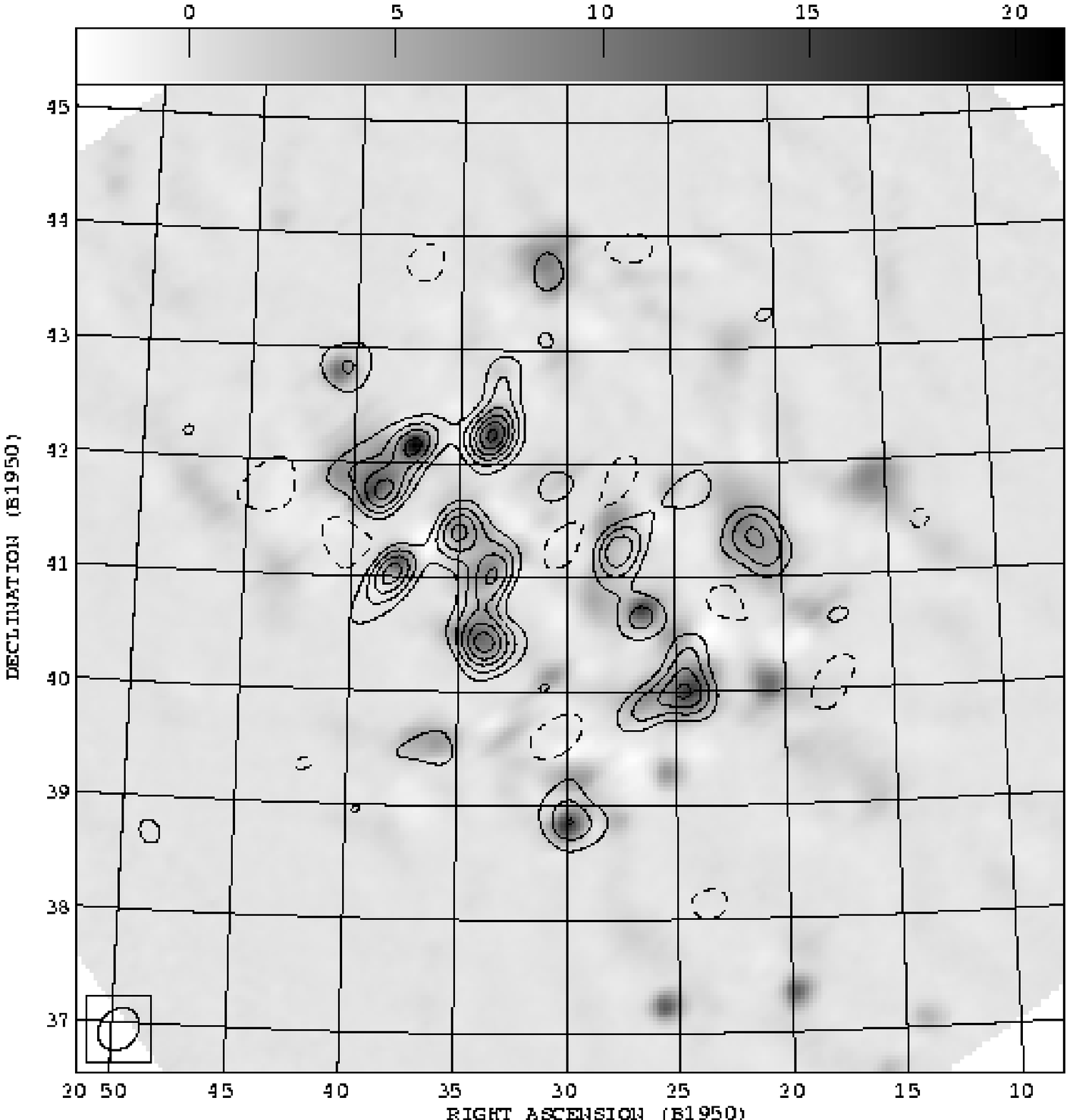}
  \caption{(a) An 80-minute observation of Jupiter, contoured at
  0.3~Jy. The thermal noise level is
    240~Jy~(baseline~$\times$~sec)$^{-1/2}$.
(b) Commissioning
    observation of the Cygnus Loop. Contours (at 3~Jy) are VSA
    observations at 34~GHz. The grey scale is Green Bank, 4.85~GHz
    data \cite{cyg_loop}.}
\end{figure}

\subsection{Calibration}
Absolute flux calibration of VSA observations is based on the flux
scale of Mason et al. (1999) {\citep{mason_casscal}.  Flux
  calibrations are made each day using one of three primary
  calibrators -- Tau A, Cass A and Jupiter.  We also make daily
  observations of three fainter sources, 3C48, 3C273 and NGC7207
  allowing us to check the quality of observations throughout the
  observing day.  The measured flux ratios of these sources to our
  primary calibrators agree well with those reported by Mason et
  al., suggesting that the accuracy of our flux calibration is limited
  by that of Mason et al. to approximately 5 percent.  
  
  The overall gain of the telescope is also monitored via a noise
  injection system. A modulated noise signal is injected into each
  antenna and is later measured using phase-sensitive detection after
  the automatic gain control stage of the telescope.  The relative
  contribution of the constant noise source to the total output power
  from each antenna varies inversely with system temperature, and thus
  a correction can be made to the overall flux calibration.  This
  system allows us to account for both variations in the gain of the
  system and for atmospheric attenuation of the astronomical signal.
  It provides an excellent indication of the weather conditions and is
  used as a primary indicator for flagging data. For good observing
  conditions, the gain corrections applied using this system are
  typically less than a few percent.
  
  Phase calibration of the VSA is also applied on a daily basis using
  the same three primary calibrators.  We find that the VSA is
  relatively phase stable, with variations of less than 15 degrees per
  day.  A typical calibration observation of Jupiter is shown in Fig.
  4(a). This 80-minute observation, made early on in the observing
  program, also confirms that the telescope is achieving the required
  sensitivity, with a thermal noise level of approximately 240~Jy/(baseline$\times$sec)$^{1/2}$.

  The pointing accuracy of the VSA is primarily determined by
  mechanical alignment tolerances, but we frequently make long
  observations of unresolved calibrators in order to check both the
  pointing and geometry of the array.  Using these observations, and
  in conjunction with a model of the telescope, we employ a
  maximum-likelihood technique to simultaneously fit for $\sim$400
  parameters.  These include the $x$, $y$ and $z$ co-ordinate of each
  antenna, correlator gains for each of the 182 correlator channels
  and the effective observing bandwidth of each baseline.  Further
  observations of offset sources and, for example, the Cygnus Loop,
  confirm the ability of the VSA to map known structure. Fig 4(b)
  shows the result of a 90-minute commissioning observation of the
  Cygnus Loop.  Our observed 34~GHz flux contours are overlaid on
  15~GHz Green Bank data {\cite{cyg_loop}. As shown in Fig. 4(b),
    there is good agreement between the structure observed at the two
    frequencies.

\section{Current Status and Conclusions}
 \begin{figure}[t]
   \centering
 \includegraphics[angle=-90,scale=0.4]{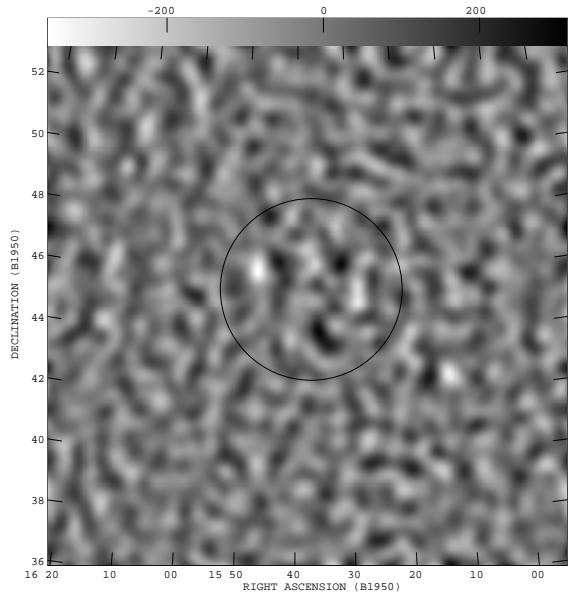}
\caption{An example of a CMB field observed with the VSA. This map is
  the result of $\sim$ 400 hours of observation. The map has had a gaussian taper applied to enhance
  the CMB features present within the FWHM (black line) of the VSA primary beam.}
\end{figure}
The VSA has been making routine observations of CMB fields in its
compact configuration since September 2000.  Commissioning and
calibration observations confirm that the instrument is working to
specification. We have currently completed over 3000 hours of
observations, having lost less than 10 percent of our observing time
due to bad weather.  A typical observation of one of our CMB fields is
shown in Fig. 5.  This map is the result of $\sim$~400 hours of
observation and, to
enhance the CMB features present in the centre, a gaussian taper has
been applied (1/e point = 60 $\lambda$). The resolution of the map is
31 $\times$ 25 arcmin. The map has not been source-subtracted.

Our observing program using the compact array was completed at the
beginning of September 2001, and we are currently re-configuring and
upgrading the array ready for observations in its extended
configuration. The second season of observations with
the new array will begin in October 2001. We will then be able to measure the CMB
power spectrum for $l$-values up to $\sim$ 1800.

\begin{theacknowledgments}
  The VSA project has involved the work of a large number of people
  from the three collaborating institutions; the Cavendish
  Astrophysics Group, Cambridge, Jodrell Bank Obsevatory, Manchester,
  and the Instituto de Astrofisica de Canarias (IAC), Tenerife.
\end{theacknowledgments}

\bibliographystyle{aipprocl} \bibliography{italy}

\end{document}